\documentclass{article}
\usepackage[super, comma, numbers, square, sort&compress]{natbib}
\usepackage{fancyvrb}
\VerbatimFootnotes 
\usepackage{setspace}
\usepackage{hyperref}
\usepackage{mciteplus}
\usepackage{lineno,hyperref}
\modulolinenumbers[5]
\usepackage{textgreek}
\usepackage{textcomp}
\usepackage{float}
\usepackage{amsmath}
\usepackage{mathtools}
\usepackage{mhchem}
\usepackage{array}
\usepackage{booktabs}
\usepackage{breqn}
\usepackage{color}
\usepackage{amssymb}
\usepackage{rotating}
\usepackage{ragged2e}
\usepackage{setspace}
\def\doubleunderline#1{\underline{\underline{#1}}}

\begin{document}

\begin{spacing}{1.75}
\center{\LARGE{A Theoretical Framework for the Electrochemical Characterization of Anisotropic Micro-Emulsions}}
\end{spacing}

\vspace*{5mm}

\center{Tim Tichter\textsuperscript{a,*}, Rohan Borah\textsuperscript{b}, Thomas Nann\textsuperscript{b}}
\vspace*{2mm}
\center{\textsuperscript{a} \scriptsize{SurfCat Section for Surface Physics and Catalysis, Department of Physics, Technical University of Denmark, 2800 Kgs Lyngby, Denmark} }
\center{\textsuperscript{b} \scriptsize{School of Mathematical and Chemical Sciences The University of Newcastle, Newcastle, NSW 2308, Australia} }
\vspace*{5mm}

\justifying
\section*{Abstract}
Micro emulsions (MEs) offer an exceptionally broad spectrum of applications covering sensing, electrosynthesis, supercapacitors and redox-flow batteries. \linebreak Herein, we develop the theory for a sophisticated electrochemical characterizion of MEs with a spatially and time-invariant anisotropy in the diffusion domain by means of cyclic voltammerty (CV). By introducing spatially dependent diffusion coefficients into Ficks´ first law, we derive a modified diffusion equation to simulate any inherent anisotropy of the ME under investigation. Moreover, by formulating an extended second-order homogeneous six-member square scheme for a kinetically controlled two-step two-electron reaction we capture the intricate entanglement of chemical and electrochemical equilibria. Our theoretical concept is finally validated by experimental CV data for the ME based two-step redox reaction of methyl-viologen which paves the way for a quantitative electrochemical characterization MEs.

\vspace*{3mm}

\vspace{5mm}
\noindent \textit{Kexwords:} \newline
Micro-Emulsions; Redox-Flow Batteries; Two-Electron Transfer; Methyl-Viologen; E\textsubscript{k}E\textsubscript{k}-Mechanism
\vspace{5mm}

\noindent \textit{Corresponding author:}\newline Tim Tichter, timtic@dtu.dk

\newpage

\section{Introduction}
Investigating electrochemical reactions with more than a single electron transfer is of utmost relevance since they occur in a plethora of chemical~\cite{Wang2015}, pharmaceutical~\cite{Hillard2008} and electrochemical~\cite{Hammerich2015, Kurniawan2016, Hu2018} frameworks. Particularly in the context of energy conversion and storage, two-electron reactions are an essential subject to study, since they can --- at least in theory --- provide twice the volumetric capacity when compared to systems with just a single electron transfer. 

Consequently, utilizing two-electron transfer sources such as viologens~\cite{Kurniawan2016, Hu2018, Janoschka2016} in the field of organic redox-flow batteries~\cite{Hammerich2015} (ORFBs) can be exceptionally beneficial to address the low energy density of the electrolytes. The major challenge is, however, the technical implementation of such two-electron transfer sources, since not all oxidation states of the depolarizer(s) are soluble in the (aqueous or organic) electrolyte to the same extent~\cite{Hu2018}. Likewise, significant precipitation might occur once a certain potential is exceeded and the electron transfer takes place. This will eventually lead to non-reversible processes\footnote{The word \textit{non-reversible} it deliberately chosen here and does --- particularly not --- refer to electrochemical reversibility or irreversibility~\cite{Matsuda1954}.} or to surface confinded reactions~\cite{Lopez2014} where the concentration of the active substance in the electrolyte fades continuously. As a consequence, one usually restricts the range of operating potentials of the RFB in a way that just one electron transfer takes place~\cite{Hu2018, Janoschka2016} --- and only 50\% of the theoretical volumetric capacity is utilized. 

\begin{figure}[H] 
\begin{center}
\includegraphics[width=6cm, height=5.4cm]{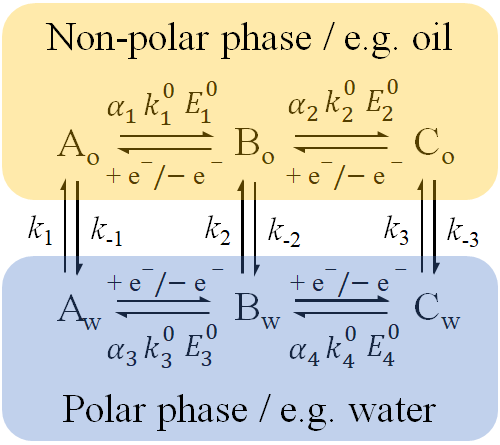}
\caption{Extended kinetic square scheme for an electrochemical E\textsubscript{k}E\textsubscript{k} reaction in a micro-emoulsion of a non-polar (oil) and a polar (water) phase. }
\label{Fig1_Reation_Scheme}
\end{center}
\end{figure}

To overcome these drawbacks, micro-emulsions (MEs), i.e. a thermodynamically stable mixture of a polar and a non-polar phase have been proposed as ORFB electrolytes recently~\cite{Peng2020}. Since MEs can simultaneously act as a polar and a non-polar solvent, they feature that different oxidation states of the depolarizer can be dissolved. Instead of precipitating, the active species formed at the electrode surface may undergo a phase transfer at a liquid/liquid interface which might be treated in turn as a homogeneous chemical reaction coupled to the charge transfer. Since any electron transfer reaction in a ME may thus occur from either phase, a successive two-step two-electron reaction might be regarded as a kinetically controlled version of the well-known extended six-member square scheme which is depicted in figure~\ref{Fig1_Reation_Scheme}. Though the theoretical model of such square scheme(s) is well established for reversible electron transfers~\cite{Lerke1990, Laborda2015, Molina2015, Molina2016, Laborda2017, Molina2018} the implementation of Butler--Volmer electrode kinetics is performed rarely~\cite{Molina2018}. Usually --- i.e. for the purpose of electrochemical trace analysis and sensing applications --- this somewhat simplified treatment is fairly sufficient, since the settings of the electroanalytical experiment can be tuned in a way such that the electrochemical kinetics can be neglected~\cite{Lopez2014}. However, this particular simplification will not be well-suited for the investigation of a ME for RFB applications, since the electrode kinetics are one essential parameter which governs the overall efficiency of a cell and which has to be meticulously regarded therefore. Apart from the electrode kinetics, another essential parameter for the accurate investigation and modeling of ME based systems, is the inherent anisotropy of the liquid bulk phase at the boundary electrode/electrolyte.

To the best of our knowledge, this factor has, to date, never been discused in the literature. Basically, a ME might be treated as a system of two immiscible phases, where the depolarizer has access to both phases --- i.e. to some extent similar to ref~\cite{Olmos2018}. However, the diffusion coefficients of the electroactive species in the respective phases might differ significantly which gives rise to the features reported in ref~\cite{Laborda2015a}. Additionally, when considering a mixture of a polar and a non-polar solvent as electrolyte, it is straightforward to envision that in direct proximity to a non-polar electrode surface (e.g. a simple glassy carbon electrode), the local concentration of the non-polar phase (and species) will be enhanced. This scenario will be to somewhat similar to the recently presented theory of a thick-film redox-layer coated electrode~\cite{Laborda2020}, however, complicated by a more or less smooth transition between non-polar and polar phases which was experimentally observed by Peng~\cite{Peng2020} from neutron scattering experiments on a hydrophobic silane modified electrode immersed into a ME electrolyte. Owing to this particular anisotropy of the liquid phase, it might be considered that the mass-transfer of the electrochemically active species in the electrolyte will become a spatially dependent quantity too. As a consequence, the diffusion coefficient has to be regarded as a function of the relative distance perpendicular to the electrode as well, which leads to the premonition that the classical diffusion is insufficient for the theoreticel treatment of MEs.

To close these essential gaps in the theoretical treatment of MEs, we develop a strategy which includes all the aforementioned quantities, i.e. 
\begin{enumerate}
    \item electrode kinetics of an E\textsubscript{k}E\textsubscript{k} reaction,
    \item spatically dependent diffusion coefficients,
    \item homogeneous chemical reactions coupled to the charge transfer step(s).
\end{enumerate}
\newpage
By considering that the inherent anisotropy of the polar to non-polar phase ratio will introduce a spatial dependence of the global diffusion coefficients\footnote{The term \textit{global diffusion coefficient} refers to the diffusion coefficient of a species in the entire electrolyte, i.e. the emulsion as medium. To illustrate this, consider a mixture of polar and a non-polar solvent. Though the diffusion coefficient of the species dissolved in one phase might not change at all, the overall diffusivity is reduced by the ratio of the volume of this particular phase to total electrolyte volume.}, we will first derive a modified version of the diffusion equation. This modified diffusion equation is subsequently solved numerically by means of digital simulation, i.e. by the Crank--Nicolson technique, for a planar, semi-infinite diffusion domain. To include the effect of electrode kinetics, the Butler--Volmer equation is used as an implicit flux-boundary at the electrode surface in all cases. The homogeneous phase transfer reactions which are coupled to the charge transfer step(s) are finally treated in terms of an extended six-member square scheme. To ultimately corroborate our theoretical model, the results from our simulations are validated against experimental data which is acquired for the electrochemical redox reactions of methyl-viologen in a water/toluene micro-emulsion. Methyl viologen serves as a well-suited example since a) its two-electron voltammetry in MEs is well-established~\cite{Mackay1990, Mackay1996} and b) the polarities of its three oxidation states are considerably different. In this manner, we gain a conclusive picture of the complex interplay of diffusive mass transfer and electrode kinetics in anisotropic electrolytes which paves the way for a decent evaluation of electrochemical experiments involving MEs. This is of utmost relevance for the recently emerging field of MEs for bulk electro-chemical processes which are not limited to ORFBs, but rather include high-voltage supercapacitors~\cite{Hughson2021}, organic electro-synthesis~\cite{Mackay1990, Mackay1996}, electro-polymerisation~\cite{Zhang2018} and bio-sensors (simultaneous detection of differently polar analytes~\cite{Kunitake2016}).

\section{Theory}
\begin{figure}[H] 
\begin{center}
\includegraphics[width=11cm, height=3.2cm]{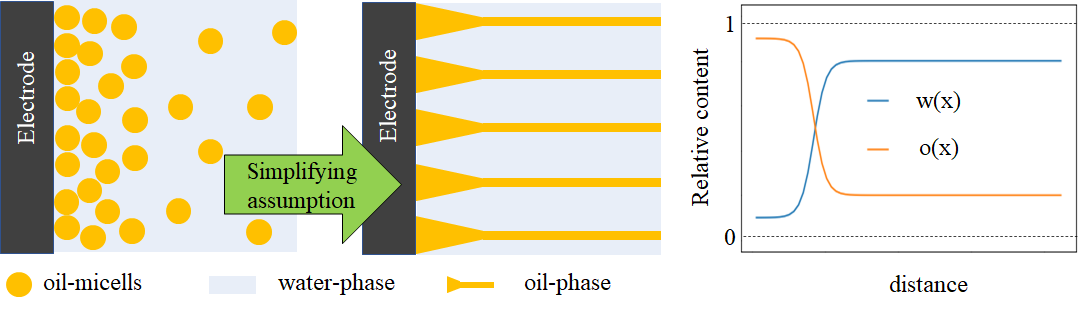}
\caption{Model for a oil in water micro-emulsion (non-polar in polar phase) with an anisotropic distribution. The oil content at the electrode is fixed at 90\%, whereas the oil content in the bulk of the electrolyte is just 20\%. Respective oil/water ratios are represented by the functions $o(x)$ and $w(x)$, which are defined by equation~\ref{o_x}.}
\label{Fig2_oil_in_water}
\end{center}
\end{figure}

The entire theory which is presented in this section is based on the assumption that the micro-emulsion under investigation is a spatially anisotropic medium, i.e. it has a time-independent non-equipartition of polar and unpolar phases. This is illustrated in figure~\ref{Fig2_oil_in_water} for a oil in water (i.e. a non-polar phase in a polar phase) micro-emulsion.

Since all electrochemically active species have access to both --- the polar and the non-polar phase --- six diffusion modes (three species times two phases) have to be considered. Likewise, the computation requires for a simultaneous solution of six homogeneously coupled concentration profiles. 

In the following, it is assumed that the global rate of diffusion of any of the species in a phase (either polar or non-polar) is restricted to the relative content of this particular phase in the electrolyte volume. Likewise, the diffusion coefficients are linearly bound to the functions $o(x)$ and $w(x)$ in figure~\ref{Fig2_oil_in_water} and are therefore considered as spatially dependent quantities. 

\newpage

\noindent To capture this effect, we introduce
\begin{equation}
D_{S,p}(x) = D_{\textrm{S,p}}^{0}p(x)
\label{Dw_equation}
\end{equation}
\noindent where $p$ denotes the phase according to $p = o,\,w$ and $S$  stands for the respective species according to $S = A,\,B,\,C$. The function $o(x)$ is defined by
\begin{equation}
o(x) =o_{\textrm{rel}}(\infty) + \dfrac{o_{\textrm{rel}}(0) - o_{\textrm{rel}}(\infty)}{1 + \textrm{exp}(\left(\dfrac{\mu [x - x_{T}]}{x_{\textrm{max}}}\right)}
\label{o_x}
\end{equation}

\noindent where $o_{\textrm{rel}}(0)$ and $o_{\textrm{rel}}(\infty)$ are the relative content of the non-polar phase at the electrode surface ($x=0$) and in the bulk of the semi-infinite\footnote{Later on, in the simulation, there is no true semi-infinite electrolyte. Instead, three times the diffusion length is considered as size of the diffusion domain in order to introduce a quasi-semi-infinite diffusion.} electrolyte ($x = \infty$). The parameter $\mu$ in equation~\ref{o_x} denotes the steepness of the polarity change when $x$ is progressively increased to $x_{\textrm{max}}$, $x_{\textrm{T}}$ is the relative twisting point in the ME profile and $x_{\textrm{max}}$ is the largest spatial distance from the electrode surface required to ensure (quasi)-semi-infinite diffusion\footnote{Quasi-semi-infinite implies that the concentration profile at the spatial boundary opposite to the electrode is affected by less than one percent and thus semi-infiniteness can be assumed.}. Once $o(x)$ is defined, $w(x)$ is simply $w(x) = 1 - o(x)$ which defines the spatically dependent diffusion coefficients as well.
\newpage 

\subsection{Spatially Dependent Diffusion Coefficients}

\noindent Since spatially dependent diffusion coefficients will inevitably lead to a more complicated mass transfer, the diffusion equation has to be rewritten. First, consider Ficks´ first law with a spatially dependent diffusion coefficient as
\begin{equation}
J(x) = -D(x)\dfrac{\partial c(x,t)}{\partial x}
\label{Fick_I_x}
\end{equation}

\noindent Noting that the time-dependent change in the concentration in a given volume element $\textrm{d}V = A\,\textrm{d}x$ --- i.e. the difference in the amount of active species entering the volume element at $x$ and leaving the volume element at $x + \textrm{d}x$ ---  can be written as
\begin{equation}
\dfrac{\partial c(x,t}{\partial t} = \dfrac{A[J(x) - J(x + \textrm{d}x)]}{\textrm{d}V} = -\dfrac{\partial J(x)}{\partial x}.
\label{Fick_II_x_deri}
\end{equation}

\noindent The differential notation in equation~\ref{Fick_II_x_deri} can be introduced since\linebreak $J(x) - J(x + \textrm{d}x) = -\textrm{d}J(x)$. Thus, perfoming the spatial derivative of the flux which is given in equation~\ref{Fick_II_x_deri} on equation~\ref{Fick_I_x} gives a modified version of the diffusion equation according to
\begin{equation}
\dfrac{\partial c(x,t)}{\partial t} = D(x)\dfrac{\partial^{2}c(x,t)}{\partial x^{2}} + \dfrac{\partial D(x)}{\partial x}\dfrac{\partial c(x,t)}{\partial x}.
\label{Fick_II_x_mod}
\end{equation}

\noindent It is readily seen from equation~\ref{Fick_II_x_mod}, that in case of a spatially independent diffusion coefficient (i.e. $D(x) = D$), the classical diffusion equation will be obtained since the second summand on the right hand side will be zero. However, since we will require spatially dependent diffusion coefficients, equation~\ref{Fick_II_x_mod} cannot be simplified.

\subsection{Defining the Diffusion-Reaction Equations}

Subsequently to including spatially dependent diffusion coefficients in Ficks´ second law, the homogeneous chemical equilibria accounting for the phase transfer reactions inside of the electrolyte have to be considered. For species $S = A,\,B,\,C$, we consider the following second-order homogeneous chemical reaction
\begin{align}
\ce{ $S_{o}(x)$ + $W(x)$ <=>[\text{\protect{\normalsize $k_{i}$}}][\text{\protect{\normalsize $k_{-i}$}}] $S_{w}(x)$ + $O(x)$},
\label{PrecedEq}
\end{align}
where the index $i = 1,\,2,\,3$ stands for the rate constants depicted in figure~\ref{Fig1_Reation_Scheme}. $O(x)$ and $W(x)$ are the absolute concentrations of the non-polar and the polar phase (in contrast to $o(x)$ and $w(x)$, which are the relative concentrations). This leads to the following general expressions for the concentration of species $S = A,\,B,\,C$ in the non-polar and polar phases, respectively. For the non-polar phase (index $o$ for oil), we have 
\begin{multline}
\dfrac{\partial S_{o}(x,t)}{\partial t} = D_{S,o}(x)\dfrac{\partial^{2}S_{o}(x,t)}{\partial x^{2}} + \dfrac{\partial D_{S,o}(x)}{\partial x}\dfrac{\partial S_{o}(x,t)}{\partial x} \\[2mm]+ k_{i}O(x)S_{w}(x,t) - k_{-i}W(x)S_{o}(x,t)  .
\label{Fick_II_x_mod_oil}
\end{multline}

\noindent In analogy, the diffusion-reaction equation for the species in the polar phase (index $w$ for water) is found as
\begin{multline}
\dfrac{\partial S_{w}(x,t)}{\partial t} = D_{S,w}(x)\dfrac{\partial^{2}S_{w}(x,t)}{\partial x^{2}} + \dfrac{\partial D_{S,w}(x)}{\partial x}\dfrac{\partial S_{w}(x,t)}{\partial x} \\[2mm]- k_{i}O(x)S_{w}(x,t) + k_{-i}W(x)S_{o}(x,t)  .
\label{Fick_II_x_mod_wat}
\end{multline}

\subsection{Solving the Diffusion-Reaction Equations}

Solving equations~\ref{Fick_II_x_mod_oil} and~\ref{Fick_II_x_mod_wat} is performed numerically by means of the Crank-Nicolson (CN) technique~\cite{Crank1996} --- a popular mathematical tool for electrochemical simulations introduced by Heinze and Störzbach~\cite{Heinze1984, Stoerzbach1993}. The (semi)implicit nature of the Crank-Nicolson method features that it provides unconditional stability which is of utmost importance for solving the coupled chemical equilibria\footnote{Though well known in electrochemistry~\cite{Magno1982, Seeber1981, Feldberg1990}, explicit methods will face significant problems in case of large homogeneous rate constants.}. The CN technique is based on an approximation of the differential notation in equations~\ref{Fick_II_x_mod_oil} and~\ref{Fick_II_x_mod_wat} by finite differences. It is usually termed a \textit{semi-implicit} technique since it approximates the spatial-derivatives as an avarage of the known (old) and yet unknown (new) concentration values which results in second order (instead of just first order) accuracy in the time coordinate. 
Setting $S_{o}(x,t+\Delta t) = S'_{o,x}$ and $S_{o}(x,t) = S_{o,x}$, the CN discretization of equation~\ref{Fick_II_x_mod_oil} is 
\begin{multline}
\dfrac{S'_{o,x}-S_{o,x}}{\Delta t} =\\[3mm] \dfrac{D_{S,o,x}}{2}\left[\dfrac{S_{o,x+\Delta x} - 2S_{o,x} + S_{o,x-\Delta x}}{\Delta x^{2}} + \dfrac{S'_{o,x+\Delta x} - 2S'_{o,x} + S'_{o,x-\Delta x}}{\Delta x^{2}}\right] \\[3mm] + \dfrac{D_{S,o,x+\Delta x} - D_{S,o,x}}{2\Delta x}\, \dfrac{1}{2}\left[\dfrac{S_{o,x+\Delta x} - S_{o,x-\Delta x}}{2\Delta x} + \dfrac{S'_{o,x+\Delta x} - S'_{o,x-\Delta x}}{2\Delta x}\right]\\[3mm] + k_{i}O_{x}\dfrac{S'_{w,x} + S_{w,x}}{2} - k_{-i}W_{x}\dfrac{S'_{o,x}+S_{o,x}}{2}.
\label{Fick_II_x_mod_oil_FD}
\end{multline}

\noindent Equation~\ref{Fick_II_x_mod_oil_FD} has the advantage that all derivatives are of second order accuracy. For this purpose, the first spatial derivative of the diffusion coefficients and of the concentrations in equation~\ref{Fick_II_x_mod_oil} were replaced by their central three-point finite difference approximations~\cite{Britz2016}. The terms containing the homogeneous chemical reactions were also treated in terms of the CN modification, i.e. the time-average between old and new time instances was introduced. A similar discretization of equation~\ref{Fick_II_x_mod_wat}, which is performed in an analogue manner gives
\begin{multline}
\dfrac{S'_{w,x}-S_{w,x}}{\Delta t} =\\[3mm] \dfrac{D_{S,w,x}}{2}\left[\dfrac{S_{w,x+\Delta x} - 2S_{w,x} + S_{w,x-\Delta x}}{\Delta x^{2}} + \dfrac{S'_{w,x+\Delta x} - 2S'_{w,x} + S'_{w,x-\Delta x}}{\Delta x^{2}}\right] \\[3mm] + \dfrac{D_{S,w,x+\Delta x} - D_{S,w,x}}{2\Delta x}\, \dfrac{1}{2}\left[\dfrac{S_{w,x+\Delta x} - S_{w,x-\Delta x}}{2\Delta x} + \dfrac{S'_{w,x+\Delta x} - S'_{w,x-\Delta x}}{2\Delta x}\right]\\[3mm] - k_{i}O_{x}\dfrac{S'_{w,x} + S_{w,x}}{2} + k_{-i}W_{x}\dfrac{S'_{o,x}+S_{o,x}}{2}.
\label{Fick_II_x_mod_wat_FD}
\end{multline}

\noindent Subsequently to discretizing equations~\ref{Fick_II_x_mod_oil} and~\ref{Fick_II_x_mod_wat}, equations~\ref{Fick_II_x_mod_oil_FD} and~\ref{Fick_II_x_mod_wat_FD} can be separated into old (known) and new (yet unknown) concentration values. Before rearrangement, the dimension-less parameter $\lambda_{S,p,x} =D_{S,p,x}\Delta t/\Delta x^{2} $ is introduced where again $p$ denotes the phase and $S$ the species. Furthermore --- and for the sake of simplicity --- the counting of spatial gridpoints is introduced in a way that $x$ at the electrode surface is defined as $x = 0$. Then, $x + \Delta x = 1$, $x + 2\Delta x = 2$ and $x + n\Delta x = n$. For the point $1$, equations~\ref{Fick_II_x_mod_oil_FD} and~\ref{Fick_II_x_mod_wat_FD} become
\begin{multline}
-[\lambda_{S,o,0} + 4 \lambda_{S,o,1} - \lambda_{S,o,2}]S'_{o,0}  +  4[2+2\lambda_{S,o,1} + \Delta t\,k_{-i}W_{1}]S'_{o,1} \\[1mm]  -[-\lambda_{S,o,0} + 4 \lambda_{S,o,1} + \lambda_{S,o,2}]S'_{o,2} - 4\Delta t\,k_{i}O_{1}S'_{w,1}  = \\[1mm] [\lambda_{S,o,0} + 4\lambda_{S,o,1} - \lambda_{S,o,2}]S_{o,0} +
4[2 - 2\lambda_{S,o,1} - \Delta t\,k_{-i}W_{1}]S_{o,1}\\[1mm] +[-\lambda_{S,o,0} + 4\lambda_{S,o,1} + \lambda_{S,o,2} ]S_{o,2} + 4\Delta t\,k_{i}O_{1}S_{w,1}
\label{Fick_II_x_mod_oil_FD_sorted}
\end{multline}

\begin{multline}
-[\lambda_{S,w,0} + 4 \lambda_{S,w,1} - \lambda_{S,w,2}]S'_{w,0}  +  4[2+2\lambda_{S,w,1} + \Delta t\,k_{i}O_{1}]S'_{w,1} \\[1mm]  -[-\lambda_{S,w,0} + 4 \lambda_{S,w,1} + \lambda_{S,w,2}]S'_{w,2} - 4\Delta t\,k_{-i}W_{1}S'_{o,1}  = \\[1mm] [\lambda_{S,w,0} + 4\lambda_{S,w,1} - \lambda_{S,w,2}]S_{w,0} +
4[2 - 2\lambda_{S,o,1} - \Delta t\,k_{i}O_{1}]S_{w,1}\\[1mm] +[-\lambda_{S,w,0} + 4\lambda_{S,w,1} + \lambda_{S,w,2} ]S_{w,2} + 4\Delta t\,k_{-i}W_{1}S_{o,1},
\label{Fick_II_x_mod_wat_FD_sorted}
\end{multline}

\noindent respectively. In equations~\ref{Fick_II_x_mod_oil_FD_sorted} and~\ref{Fick_II_x_mod_wat_FD_sorted} it should be noted that all the coefficients $\lambda$ are known if $\Delta x$ and $\Delta t$ are given, since the spatial dependence of the diffusion coefficients is provided by equation~\ref{Dw_equation}. Likewise, all terms in the square brackes in equations~\ref{Fick_II_x_mod_oil_FD_sorted} and~\ref{Fick_II_x_mod_wat_FD_sorted} might be re-defined for the sake of clarity. Putting
\begin{equation}
\theta_{S,p,1} = \lambda_{S,p,0} + 4 \lambda_{S,p,1} - \lambda_{S,p,2}
\label{epsilons}
\end{equation} 
\begin{equation}
\sigma_{S,p,1} = -\lambda_{S,p,0} + 4 \lambda_{S,p,1} + \lambda_{S,p,2}
\label{sigmas}
\end{equation} 
\begin{equation}
\psi_{S,p,1} = 4[2-2\lambda_{S,p,1}]
\label{psi}
\end{equation}
\begin{equation}
\psi '_{S,p,1} = 4[2+2\lambda_{S,p,1}]
\label{psiprime}
\end{equation}
\begin{equation}
\phi_{\pm i,p,1} = 4\Delta t\,k_{\pm i}P_{1}
\label{phi}
\end{equation}
\noindent we have
\begin{multline}
-\theta_{S,o,1}S'_{o,0}  +  [\psi '_{S,o,1} + \phi_{-i,w,1} ]S'_{o,1} - \sigma_{S,o,1}S'_{o,2} - \phi_{i,o,1}S'_{w,1}   
 \\[-0.5mm]=\\[-0.5mm] \theta_{S,o,1}S_{o,0} +
[\psi_{S,o,1} - \phi_{-i,w,1}]S_{o,1} + \sigma_{S,o,1}S_{o,2} + \phi_{i,o,1}S_{w,1}
\label{Fick_II_x_mod_oil_FD_sorted_short}
\end{multline}

\noindent and
\begin{multline}
-\theta_{S,w,1}S'_{w,0}  +  [\psi '_{S,w,1} + \phi_{i,o,1} ]S'_{w,1} - \sigma_{S,w,1}S'_{w,2} - \phi_{-i,w,1}S'_{o,1}   
 \\[-0.5mm]=\\[-0.5mm] \theta_{S,w,1}S_{w,0} +
[\psi_{S,w,1} - \phi_{i,o,1}]S_{w,1} + \sigma_{S,w,1}S_{w,2} + \phi_{-i,w,1}S_{o,1}.
\label{Fick_II_x_mod_wat_FD_sorted_short}
\end{multline}

\noindent Equations~\ref{Fick_II_x_mod_oil_FD_sorted_short} and~\ref{Fick_II_x_mod_wat_FD_sorted_short} are valid for all species and all phases involved, by simply adjusting the indices. However, they were derived for the first spatial gridpoint only. The other gridpoints can be accessed in the same way by successively increasing the numerical index by integer values. To account for all spatial gridpoints simultaneously, equations~\ref{Fick_II_x_mod_oil_FD_sorted_short} and~\ref{Fick_II_x_mod_wat_FD_sorted_short} are specified for their respective phases and arragned into a matrix notation. For this purpose it is, however, of utmost importance to notice that  ---  for one species ($A,\,B,\,C$) --- both equations are coupled. Furthermore, since all species are linked by the electrochemical reactions at the electrode surface, all concentration profiles need to be solved simultaneously. Define for an exemplary spatial grid consisting of $N = 7$ grid points counting from 0 to 6 where the zeroth gridpoint is not taken into the matrix

\begin{equation}
\doubleunderline{M}_{S,p} = 
\begin{pmatrix}
\psi_{S,p,1}    & \sigma_{S,p,1} &  & &  \\
\theta_{S,p,2} & \psi_{S,p,2}  & \sigma_{S,p,2} & &  \\
 & \theta_{S,p,3} & \psi_{S,p,3}  & \sigma_{S,p,3} &  \\
 &   & \theta_{S,p,4} & \psi_{S,p,4}  & \sigma_{S,p,4} \\
 &  &   & \theta_{S,p,5} & \psi_{S,p,5} \\
\end{pmatrix}
\label{M_old}
\end{equation}

\begin{equation}
\doubleunderline{M'}_{S,p} = 
\begin{pmatrix}
\psi '_{S,p,1}    & -\sigma_{S,p,1} &  & &  \\
-\theta_{S,p,2} & \psi '_{S,p,2}  & -\sigma_{S,p,2} & &  \\
 & -\theta_{S,p,3} & \psi '_{S,p,3}  & -\sigma_{S,p,3} &  \\
 &   & -\theta_{S,p,4} & \psi '_{S,p,4}  & -\sigma_{S,p,4} \\
 &  &   & -\theta_{S,p,5} & \psi '_{S,p,5} \\
\end{pmatrix}
\label{M_new}
\end{equation}

\begin{equation}
\doubleunderline{\Phi}_{\pm i,p} = 
\begin{pmatrix}
\phi_{\pm i,p,1} & & & &  \\
 &\phi_{\pm i,p,2} & & &  \\
 & &\phi_{\pm i,p,3} & &  \\
 & & &\phi_{\pm i,p,4} &  \\
 & & & &\phi_{\pm i,p,5}  \\
\end{pmatrix}
\end{equation}

\noindent Then, the diffusion part of concentration profiles related to the reaction scheme in figure~\ref{Fig1_Reation_Scheme} can be written as
\begin{equation}
\doubleunderline{\mathfrak{D}} = 
\begin{pmatrix}
\doubleunderline{M}_{A,o} &  & & & & \\
 & \doubleunderline{M}_{B,o} &  &  &  &  \\
 &  & \doubleunderline{M}_{C,o} &  &  &  \\
 &  &  & \doubleunderline{M}_{A,w} &  &  \\
 &  &  &  & \doubleunderline{M}_{B,w} &  \\
 &  &  &  &  & \doubleunderline{M}_{C,w} 
\end{pmatrix}
\label{Diffmat_old}
\end{equation}
\begin{equation}
\doubleunderline{\mathfrak{D}}' = 
\begin{pmatrix}
\doubleunderline{M}'_{A,o} &  & & & & \\
 & \doubleunderline{M}'_{B,o} &  &  &  &  \\
 &  & \doubleunderline{M}'_{C,o} &  &  &  \\
 &  &  & \doubleunderline{M}'_{A,w} &  &  \\
 &  &  &  & \doubleunderline{M}'_{B,w} &  \\
 &  &  &  &  & \doubleunderline{M}'_{C,w} 
\end{pmatrix}
\label{Diffmat_new}
\end{equation}

and the homogeneous kinetics part as

\begin{equation}
\doubleunderline{\mathfrak{K}} = 
\begin{pmatrix}
\doubleunderline{\Phi}_{-1,w} &  &  & -\doubleunderline{\Phi}_{1,o} &  &  \\
 & \doubleunderline{\Phi}_{-2,w} &  &  & -\doubleunderline{\Phi}_{2,w} &  \\
 &  & \doubleunderline{\Phi}_{-3,w} &  &  & -\doubleunderline{\Phi}_{3,w} \\
-\doubleunderline{\Phi}_{-1,w} &  &  & \doubleunderline{\Phi}_{1,o} &  &  \\
 & -\doubleunderline{\Phi}_{-2,w} &  &  & \doubleunderline{\Phi}_{2,o} &  \\
 &  & -\doubleunderline{\Phi}_{-3,w} &  &  & \doubleunderline{\Phi}_{3,o} 
\end{pmatrix}.
\end{equation}

\noindent The old (known) and new (yet unknown) concentration profile of each species $S$ in phase $p$ can be written in terms of a verctor with $N-2$ entries, if an $N$-dimensional spacegrid is used. It follows
\begin{equation}
\underline{v}_{S,p} = (S_{1,p}, S_{2,p}, ... , S_{N-3,p},S_{N-2,p} )
\label{generalvect_old}
\end{equation}
\begin{equation}
\underline{v}'_{S,p} = (S'_{1,p}, S'_{2,p}, ... , S'_{N-3,p},S'_{N-2,p} )
\label{generalvect_new}
\end{equation}
Since neither equations~\ref{M_old} and~\ref{M_new}, nor equations~\ref{Diffmat_old} to~\ref{generalvect_new} account for the spatial grid point at zero (at the electrode surface) and at \textit{pseudo-infinity} (at the last spatial grid point) additional defining equations for adding the electrochemical boundary and the constant concentration boundary are required. They are given by the following vectors of dimension $N-2$
\begin{equation}
\underline{b}_{S,p} = (\theta_{S,p,1}S_{0,p},\, 0,\,...\,,\,0,\,0)
\label{boundvect_old}
\end{equation}
\begin{equation}
\underline{b}'_{S,p} = (-\theta_{S,p,1}S'_{0,p},\,0,\,...\,,\,0,\,0).
\label{boundvect_new}
\end{equation}

\begin{equation}
\underline{h}_{S,p} = (0,\, 0,\,...\,,\,0,\,\sigma_{S,p,6}S_{6,p})
\label{boundvect_old_Out}
\end{equation}
\begin{equation}
\underline{h}'_{S,p} = (0,\,0,\,...\,,\,0,\,-\sigma_{S,p,6}S'_{6,p}).
\label{boundvect_new_Out}
\end{equation}

\noindent This allows for the combination of all concentration profiles and all boundary conditions (only shown for the knowns here since the unknows are obtained similarly) as
\begin{equation}
\underline{\mathfrak{v}} = (\underline{v}_{A,o}, \underline{v}_{B,o}, \underline{v}_{C,o}, \underline{v}_{A,w}, \underline{v}_{B,w}, \underline{v}_{C,w} )
\label{Stacked_vect_old}
\end{equation}
\begin{equation}
\underline{\mathfrak{b}} = (\underline{b}_{A,o}, \underline{b}_{B,o}, \underline{b}_{C,o}, \underline{b}_{A,w}, \underline{b}_{B,w}, \underline{b}_{C,w} ).
\label{Stacked_bound_old}
\end{equation}
\begin{equation}
\underline{\mathfrak{h}} = (\underline{h}_{A,o}, \underline{h}_{B,o}, \underline{h}_{C,o}, \underline{h}_{A,w}, \underline{h}_{B,w}, \underline{h}_{C,w} ).
\label{Stacked_bound_Out}
\end{equation}

\noindent Finally, the following matrix notation is obtained which has to be solved for each and every time interation
\begin{equation}
(\doubleunderline{\mathfrak{D}} - \doubleunderline{\mathfrak{K}})\,\underline{\mathfrak{v}} + \underline{\mathfrak{b}} + 2\,\underline{\mathfrak{h}} = \underline{u} =  (\doubleunderline{\mathfrak{D}'} + \doubleunderline{\mathfrak{K}})\,\underline{\mathfrak{v}'} + \underline{\mathfrak{b}'}.
\label{Final_System}
\end{equation}

\noindent In equation~\ref{Final_System}, $\underline{u}$ is an auxiliary result between the two timesteps. Since the concentration at the outer boundary is assumed to be constant we note that, $\underline{\mathfrak{h}} = -\underline{\mathfrak{h}}'$ such that the boundary verctor of the new time instance is known and can be added readily to the old expressions. The next step is then to implement the Butler--Volmer electrode kinetics for the two parallel E\textsubscript{k}E\textsubscript{k} reactions.

\newpage

\subsection{Implementing Butler--Volmer Electrode Kinetics}

\noindent The Butler--Volmer electrode kinetics can be included as follows. First, consider the fluxes (current normalized by Faradays´law) according to the Butler--Volmer equation 
\begin{equation}
J_{1} =  k_{f,1}A_{o,0} -k_{b,1}B_{o,0}
\label{BV_Flux_1}
\end{equation}
\begin{equation}
J_{2} = k_{f,2}B_{o,0} -k_{b,2}C_{o,0}
\label{BV_Flux_2}
\end{equation}
\begin{equation}
J_{3} = k_{f,3}A_{w,0} -k_{b,3}B_{w,0}
\label{BV_Flux_3}
\end{equation}
\begin{equation}
J_{4} = k_{f,4}A_{w,0} -k_{b,4}B_{w,0}
\label{BV_Flux_4}
\end{equation}
where
\begin{equation}
k_{f,i} = k^{0}_{i}\textrm{e}^{\alpha_{i}\xi_{i}} \hspace*{20mm} k_{b,i} = k^{0}_{i}\textrm{e}^{-(1-\alpha_{i})\xi_{i}}
\label{kfi}
\end{equation}

\noindent where $\xi_{i} = nF[E(t) - E^{0}_{i}]/RT$. Since equations~\ref{BV_Flux_1} to~\ref{BV_Flux_4} only contain the surface concentrations at $x=0$ which are not included in the matrix notations of equation~\ref{Final_System} (i.e. they are isolated in the boundary conditions) another defining equation which eliminates the surface concentrations is required. This expression is given in terms of the modified Ficks´ first law, i.e. equation~\ref{Fick_I_x}. For the non-polar phase this gives
\begin{equation}
D_{A,o,0}\dfrac{\partial A_{o}}{\partial x}\Bigg|_{x=0} = k_{f,1}A_{o,0} -k_{b,1}B_{o,0}
\label{Fick_Flux_A}
\end{equation}
\begin{equation}
D_{B,o,0}\dfrac{\partial B_{o}}{\partial x}\Bigg|_{x=0} = -k_{f,1}A_{o,0} + [k_{b,1} + k_{f,2}B_{o,0}] - k_{b,2}C_{o,0}  
\label{Fick_Flux_B}
\end{equation}
\begin{equation}
D_{C,o,0}\dfrac{\partial C_{o}}{\partial x}\Bigg|_{x=0} = - k_{f,2}B_{o,0} + k_{b,2}C_{o,0}
\label{Fick_Flux_C}
\end{equation}
and the expressions for the polar phase can be derived in the same way (changing index $o$ to $w$ and $1,2$ to $3,4$). Equations~\ref{Fick_Flux_A} to~\ref{Fick_Flux_C} (as well as their analogues for the polar phase) will be approximated in terms of finite differences. 

\noindent Since all expressions which are stated in terms of finite differences so far possess a second order accuracy, we use an asymmetric three-point forward finite difference approximation (which is of second order accuracy in space as well) at the electrode surface. For species $S$ in phase $p$, we have
\begin{equation}
\dfrac{\partial S_{p}}{\partial x}\Bigg|_{x=0} = \dfrac{-3S_{p,0} + 4S_{p,1} - S_{p,2} }{2\Delta x}.
\label{Three_Point_Approx}
\end{equation}

\noindent Introducing equation~\ref{Three_Point_Approx} in quations~\ref{BV_Flux_1} to~\ref{Fick_Flux_C} and combining the result with equations~\ref{BV_Flux_1} to~\ref{BV_Flux_4} allows to eliminate the surface concentrations. Generalizing for phase $p$, we find
\begin{equation}
A_{p,0} = aa_{p,1}A_{p,1} + aa_{p,2}A_{p,2} + ab_{p,1}B_{p,1} + ab_{p,2}B_{p,2} + ac_{p,1}C_{p,1} + ac_{p,2}C_{p,2}
\label{A_p_0}
\end{equation}
\begin{equation}
B_{p,0} = ba_{p,1}A_{p,1} + ba_{p,2}A_{p,2} + bb_{p,1}B_{p,1} + bb_{p,2}B_{p,2} + bc_{p,1}C_{p,1} + bc_{p,2}C_{p,2}
\label{B_p_0}
\end{equation}
\begin{equation}
C_{p,0} = ca_{p,1}A_{p,1} + ca_{p,2}A_{p,2} + cb_{p,1}B_{p,1} + cb_{p,2}B_{p,2} + cc_{p,1}C_{p,1} + cc_{p,2}C_{p,2},
\label{C_p_0}
\end{equation}

\noindent where the coefficients in equations~\ref{A_p_0} to~\ref{C_p_0}, were derived in this paper and are given in table~\ref{tab:aabbccCOEFFICIENTS}. The $z$-values in table~\ref{tab:aabbccCOEFFICIENTS} are defined in table~\ref{tab:zCOEFFICIENTS}.

\begin{table}[H]
\begin{center}
\caption{\label{tab:zCOEFFICIENTS} Definition of the $z$-parameters utilized in table~\ref{tab:aabbccCOEFFICIENTS} \vspace*{5mm} }
{\renewcommand{\arraystretch}{2}
\begin{tabular}{|l|l|}
\cline{1-2}
$z_{1,u} = \dfrac{1}{k^{0}_{u}\textrm{e}^{\alpha_{u}\xi_{u}}} + \dfrac{2\Delta x}{3D_{A,p,0}}$ & $z_{5,u} =z_{1,u} + \dfrac{2\Delta x \textrm{e}^{-\xi_{u}}}{3D_{B,p,0}} $ \\ \cline{1-2}
$z_{2,u} = \dfrac{\textrm{e}^{-\xi_{u}}}{z_{1,u}} + \dfrac{3D_{B,p,0}}{2\Delta x}$ & $z_{6,u} =z_{3,u} + \dfrac{2\Delta x \textrm{e}^{-\xi_{u+1}}}{3D_{C,p,0}} $\\ \cline{1-2}
$z_{3,u} = \dfrac{1}{k^{0}_{u+1}\textrm{e}^{\alpha_{u+1}\xi_{u+1}}} + \dfrac{2\Delta x \textrm{e}^{-\xi_{u+1}}}{3D_{C,p,0}}$ & $z_{7,u} = 1-\dfrac{4\Delta x^{2}\textrm{e}^{-\xi_{u}}}{3z_{5,u}z_{6,u}D_{B,p,0}^{2}}$ \\ \cline{1-2}
$z_{4,u} = z_{3,u} + \dfrac{1}{z_{2,u}} $ & $u = 
\begin{cases}
    1 ,& \text{if } p = o\\
    3 ,& \text{if } p = w 
\end{cases}
$ \\ \cline{1-2}
\end{tabular}
}
\end{center}
\end{table}

\newpage

\begin{table}[H]
\begin{center}
\caption{\label{tab:aabbccCOEFFICIENTS} Definition of the coefficients introduced in equations~\ref{A_p_0} to~\ref{C_p_0}, which are required to express the surface concentrations as functions of the implicitly computed fluxes. \vspace*{5mm} }
{\renewcommand{\arraystretch}{2}
\begin{tabular}{|l|l|}
\cline{1-2}
$aa_{p,1} = \dfrac{4}{3} - \dfrac{8\Delta x}{9D_{A,p,0}z_{1,u}}\left(1 - \dfrac{z_{3,u}\textrm{e}^{-\xi_{u}}}{z_{1,u}z_{2,u}z_{4,u}}   \right)$ & $ba_{p,1} = \dfrac{4z_{3,u}}{3z_{1,u}z_{2,u}z_{4,u}}$      \\ \cline{1-2}
$aa_{p,2} = -\dfrac{1}{3} + \dfrac{2\Delta x}{9D_{A,p,0}z_{1,u}}\left(1 - \dfrac{z_{3,u}\textrm{e}^{-\xi_{u}}}{z_{1,u}z_{2,u}z_{4,u}}   \right)$ & $ba_{p,2} = -\dfrac{z_{3,u}}{3z_{1,u}z_{2,u}z_{4,u}}$    \\ \cline{1-2}
$ab_{p,1} = \dfrac{4D_{B,p,0}z_{3,u}\textrm{e}^{-\xi_{u}}}{3D_{A,p,0}z_{1,u}z_{2,u}z_{4,u}}$ & $bb_{p,1} = \dfrac{4D_{B,p,0}z_{3,u}}{2\Delta x z_{2,u}z_{4,u}}$ \\ \cline{1-2}
$ab_{p,2} = -\dfrac{D_{B,p,0}z_{3,u}\textrm{e}^{-\xi_{u}}}{3D_{A,p,0}z_{1,u}z_{2,u}z_{4,u}}$  & $bb_{p,2} = -\dfrac{D_{B,p,0}z_{3,u}}{2\Delta x z_{2,u}z_{4,u}}$    \\ \cline{1-2}
$ac_{p,1} = \dfrac{8\Delta x \textrm{e}^{-[\xi_{u} + \xi_{u+1}]}}{9D_{A,p,0}z_{1,u}}\left(1 - \dfrac{z_{3,u}}{z_{4,u}}   \right)$ & $bc_{p,1} = \dfrac{4\textrm{e}^{-\xi_{u+1}}}{3} \left(1 - \dfrac{z_{3,u}}{z_{4,u}} \right)$ \\ \cline{1-2}
$ac_{p,2} = -\dfrac{2\Delta x \textrm{e}^{-[\xi_{u} + \xi_{u+1}]}}{9D_{A,p,0}z_{1,u}}\left(1 - \dfrac{z_{3,u}}{z_{4,u}}   \right)$& $bc_{p,2} = -\dfrac{\textrm{e}^{-\xi_{u+1}}}{3} \left(1 - \dfrac{z_{3,u}}{z_{4,u}} \right)$ \\ \cline{1-2}
$ca_{p,1} = \dfrac{8\Delta x }{9D_{C,p,0}z_{1,u}z_{2,u}z_{4,u}}$ & $cb_{p,2} = -\dfrac{D_{B,p,0}}{3D_{C,p,0}z_{2,u}z_{4,u}}$  \\ \cline{1-2}
$ca_{p,2} = -\dfrac{2\Delta x }{9D_{C,p,0}z_{1,u}z_{2,u}z_{4,u}}$& $cc_{p,1} = \dfrac{4}{3} - \dfrac{8\Delta x \textrm{e}^{-\xi_{u+1}} }{9D_{C,p,0}z_{4,u}}$ \\ \cline{1-2}
$cb_{p,1} = \dfrac{4D_{B,p,0}}{3D_{C,p,0}z_{2,u}z_{4,u}}$ & $cc_{p,2} = -\dfrac{1}{3} + \dfrac{2\Delta x \textrm{e}^{-\xi_{u+1}} }{9D_{C,p,0}z_{4,u}}$ \\ \cline{1-2}
\end{tabular}
}
\end{center}
\end{table}

\noindent Substituting all definitions of tables~\ref{tab:zCOEFFICIENTS} and~\ref{tab:aabbccCOEFFICIENTS} into equations~\ref{A_p_0} to~\ref{C_p_0} allows to invoke the old and new boundary vectors (defined exemplary for the old values in equation~\ref{Stacked_bound_old}) into diffusion matrices~\ref{Diffmat_old} and~\ref{Diffmat_new}. This allows for computing the time dependent concentration profiles at each and every time instance. 

The Butler--Volmer fluxes of all species are finally obtained by generating the improved flux expression by the procedure introduced by Heinze~\cite{Heinze1984}. 
\newpage

\noindent In the present case of an E\textsubscript{K}E\textsubscript{K}, we find
\begin{multline}
J_{1} =  \dfrac{1}{z_{5,1}z_{7,1}}\Bigg[\dfrac{4A_{o,1}-A_{o,2}}{3} - \left(1-\dfrac{2\Delta x}{3D_{B,o,0}z_{6,1}}\right)\dfrac{\textrm{e}^{-\xi_{1}}[4B_{o,1}-B_{o,2}]}{3} \\[2mm] - \dfrac{2\Delta x \textrm{e}^{-[\xi_{1} + \xi_{2}]}[C_{1,o} - C_{2,o}]}{9z_{6,1}D_{B,o,0}}     \Bigg]
\label{Final_Flux_1}
\end{multline}
\begin{equation}
J_{2} = \dfrac{1}{z_{6,1}} \Bigg[\dfrac{4B_{o,1}-B_{o,2}}{3} - \dfrac{\textrm{e}^{-\xi_{2}}[4C_{o,1} - C_{o,2}]}{3} + \dfrac{2\Delta x}{3D_{B,o,0}}J_{1}   \Bigg]
\label{Final_Flux_2}
\end{equation}
\begin{multline}
J_{3} =  \dfrac{1}{z_{5,3}z_{7,3}}\Bigg[\dfrac{4A_{w,1}-A_{w,2}}{3} - \left(1-\dfrac{2\Delta x}{3D_{B,w,0}z_{6,3}}\right)\dfrac{\textrm{e}^{-\xi_{3}}[4B_{w,1}-B_{w,2}]}{3} \\[2mm] - \dfrac{2\Delta x \textrm{e}^{-[\xi_{3} + \xi_{4}]}[C_{1,w} - C_{2,w}]}{9z_{6,3}D_{B,w,0}}     \Bigg]
\label{Final_Flux_3}
\end{multline}
\begin{equation}
J_{4} = \dfrac{1}{z_{6,3}} \Bigg[\dfrac{4B_{w,1}-B_{w,2}}{3} - \dfrac{\textrm{e}^{-\xi_{4}}[4C_{w,1} - C_{w,2}]}{3} + \dfrac{2\Delta x}{3D_{B,w,0}}J_{3}   \Bigg]
\label{Final_Flux_4}
\end{equation}

\noindent The total electric current is finally generated by 
\begin{equation}
I = nFA[J_{1} + J_{2} + J_{3} + J_{4}]
\label{CURRENT}
\end{equation}

\noindent and the dimension-less flux (similar to the Randles--\v{S}ev\v{c}ík equation) by 
\begin{equation}
\chi = [J_{1} + J_{2} + J_{3} + J_{4}]\sqrt{\dfrac{RT}{D_{A,p}^{0}F\nu}}.
\label{CHI}
\end{equation}

\noindent This allows the simulation of electroanalytical experiments with any kind of potenital program (of course including CV). The most crucial part is then the proper choice of the increments $\Delta t$ and $\Delta x$ such that the numerically computed results are sufficiently accurate. Despite the fact that owing to the unconditional stability of the Crank--Nicolson technique this restriction is usually not an issue, we fixed the maximum value of the dimension-less parameter $\lambda$ \linebreak to $\lambda_{S,p,x,\textrm{max}} = 0.5$. The simulation was then performed on a time-grid, defined by the dimension-less potential substep of $\Delta \xi = 0.1$, which can be converted into $\Delta t$-units in a straightforward way.  This finally defines the spatial increment $\Delta x$ as well as the amount of spatial gridpoints.

\newpage

\section{Results and discussion}
For all the following simulations it was assumed that only species $A_{p}$ is initially present in either phase of the electrolyte. This corresponds to an initial equilibrium potential chosen sufficiently far away from all $E^{0}$. The next step was then to compute the initial equilibrium distribution of $A_{o}$ and $A_{w}$. For this purpose, the following system of equations was solved until time-invariant concentration profiles were obtained. This introduces the balance of diffusive mass transfer and homogeneous chemical reactions.

\begin{multline}
\begin{pmatrix}
\doubleunderline{M}_{A,o}-\doubleunderline{\Phi}_{-1,w} & \doubleunderline{\Phi}_{1,o} \\
\doubleunderline{\Phi}_{-1,w} & \doubleunderline{M}_{A,w}-\doubleunderline{\Phi}_{1,o} \\
\end{pmatrix}
\begin{pmatrix}
\underline{v}_{A,o} \\
\underline{v}_{A,w} \\
\end{pmatrix}+
\begin{pmatrix}
\underline{b}_{A,o}^{*} \\
\underline{b}_{A,w}^{*} \\
\end{pmatrix}+ 2
\begin{pmatrix}
\underline{h}_{A,o} \\
\underline{h}_{A,w} \\
\end{pmatrix}
 =\\[2mm] 
\begin{pmatrix}
\doubleunderline{M}_{A,o}'+\doubleunderline{\Phi}_{-1,w} & -\doubleunderline{\Phi}_{1,o} \\
-\doubleunderline{\Phi}_{-1,w} & \doubleunderline{M}_{A,w}'+\doubleunderline{\Phi}_{1,o} \\
\end{pmatrix}
\begin{pmatrix}
\underline{v}^{'}_{A,o} \\
\underline{v}^{'}_{A,w}  \\
\end{pmatrix}
+\begin{pmatrix}
\underline{b}_{A,o}^{*'} \\
\underline{b}_{A,w}^{*'} \\
\end{pmatrix}
\label{EQUATOR_Equation}
\end{multline}

\noindent The vectors $\underline{b}_{S,p}^{*}$ and $\underline{b}_{S,p}^{*'}$ in equation~\ref{EQUATOR_Equation} introduce a no-flux boundary at the electrode surface, since no electrochemical reactions are assumed during the initial equilibration process. They have a length of $N-2$ and are defined by $\underline{b}_{S,p}^{*} = (\theta_{S,p,1}S_{1,p},\, 0,\,...\,,\,0,\,0)$ and $\underline{b}_{S,p}^{*'} = (-\theta_{S,p,1}S_{1,p},\, 0,\,...\,,\,0,\,0)$. 

\begin{figure}[H] 
\begin{center}
\includegraphics[width=10cm, height=8.8cm]{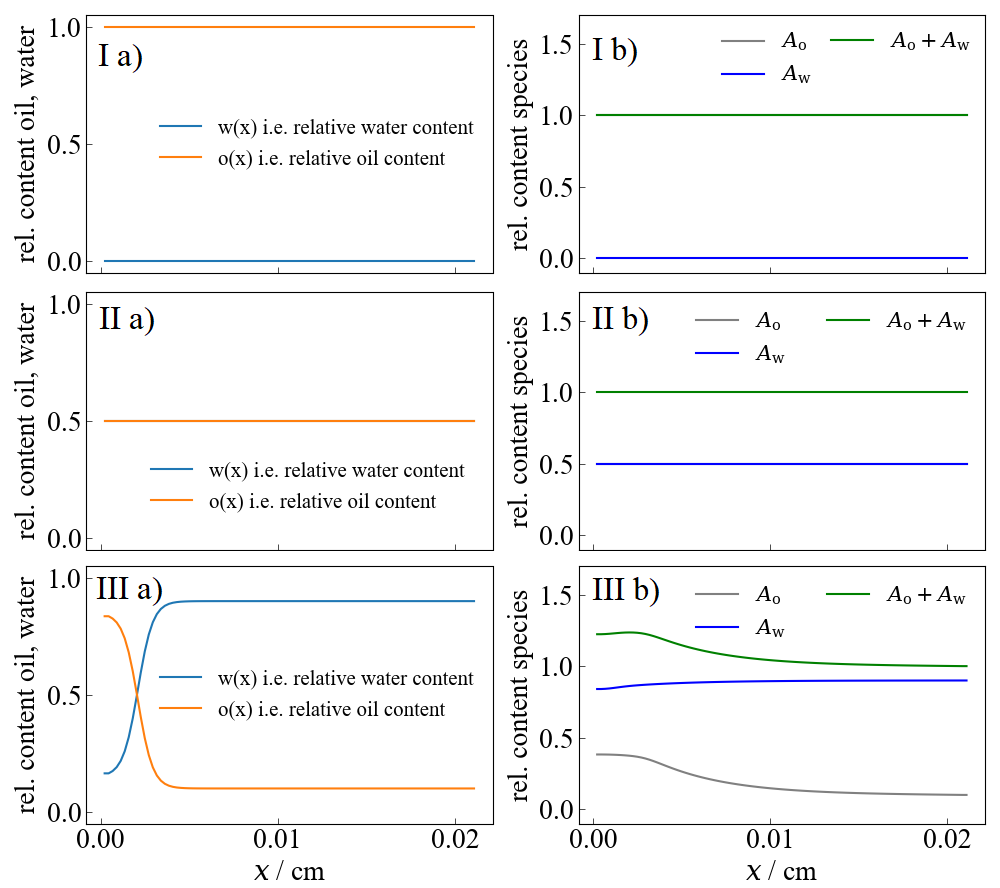}
\caption{I a), II a) III a): Functions $o(x)$ and $w(x)$ for different ratios of oil in water (I: 1/1000, II: 1/1, III: variable). I b), II b) III b): Equilibrium distribution of $A_{\textrm{o}}$ and $A_{\textrm{w}}$ according to a diffusion-reaction balance. In case of III b) it can be seen that the anisotropy in the ration of polar and non-polar phases introduces a  non-equipartition in $A_{\textrm{o}}$ and $A_{\textrm{w}}$ as well which leads to a locally increased total concentration. }
\label{Fig3_oil_in_water_And_Species}
\end{center}
\end{figure}

Figure~\ref{Fig3_oil_in_water_And_Species} depicts the spatial distributions of $A_{o}$ and $A_{w}$ for different ratios and spatial distributions of the non-polar and polar phases which were computed on the base of equation~\ref{EQUATOR_Equation} and which are the initial situations for the electrochemical reactions. It can be seen that in case of an isotropic distribution of polar and non-polar phases the species $A_{\textrm{o}}$ and $A_{\textrm{w}}$ will be equally distributed as well. However, as soon as $o(x)$ and $w(x)$ introduce an anisotropy, the equilibrium state of $A_{\textrm{o}}$ and $A_{\textrm{w}}$ will be anisotropic as well. This is demonstrated in panels III a) and III b). It can be seen that the anisotropy in $o(x)$ and $w(x)$ will even introduce a maximum in the total concentration owing to the balance of homogeneous reaction processes and diffusive mass transfer.

\begin{figure}[H] 
\begin{center}
\includegraphics[width=11cm, height=14.5cm]{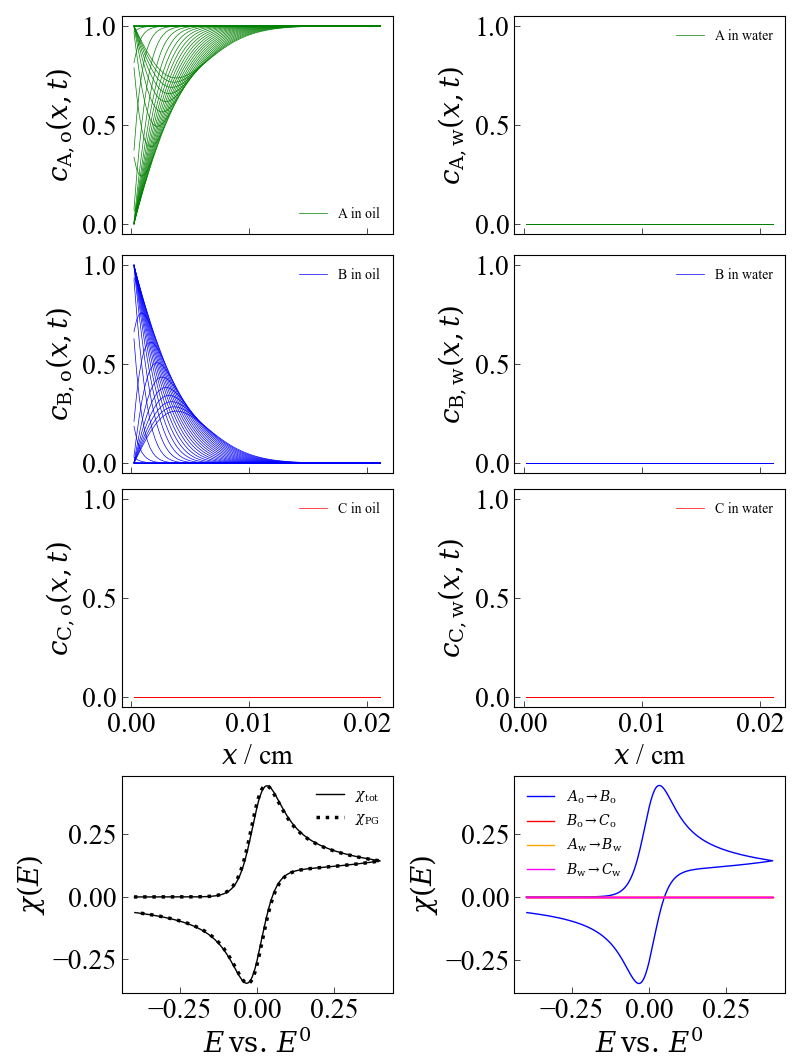}
\caption{CV curves simulated for the spatial equipartition of non-polar to polar phases in 1000/1, depicted in figure~\ref{Fig3_oil_in_water_And_Species} I a) and I b). For the simulation it was set $E^{0}_{1} = E^{0}_{3} = 0.00\,\textrm{V}$, $E^{0}_{2} = E^{0}_{4} = 0.05\,\textrm{V}$, $\alpha_{1} = \alpha_{2} = \alpha_{3} = \alpha_{4} = 0.5$, $k^{0}_{1} = k^{0}_{3} = 10^{3}\,\textrm{cm/s}$, $k^{0}_{2} = k^{0}_{4} = 10^{-23}\,\textrm{cm/s}$, $D_{A,o}^{0} = D_{B,o}^{0} = D_{C,o}^{0} = D_{A,w}^{0} = D_{B,w}^{0} = D_{C,w}^{0} = 10^{-6}\,\textrm{cm}^{2}\textrm{/s} $, $A = 1\,\textrm{cm}^{2}$, $c_{\textrm{tot}} = 1\,\textrm{mol/L}$, $k_{1} = 0.1~\textrm{mol/Ls}$, $k_{-1} = 10^{-10}~\textrm{mol/Ls}$, $k_{2} = 10^{-3}~\textrm{mol/Ls}$, $k_{3} = 10^{-3}~\textrm{mol/Ls}$ ($k_{-2}$ and $k_{-3}$ are thus defined from the thermodynamics). The function $o(x)$ was defined by setting $\mu = 50$ and $x_{\textrm{T}} = 0.5\,x_{\textrm{max}}$, $o_{\textrm{rel}}(\infty) = o_{\textrm{rel}}(0) = 0.999$. The total concentration of the polar phase was set to $55\,\textrm{mol/L}$ and the concentration of the non-polar phase to $5\,\textrm{mol/L}$.}
\label{Fig4_Classical_Comparison}
\end{center}
\end{figure}

To elucidate the influence of isotropic or anisotropic distributions of polar and non-polar phases on the electrochemistry of a micro-emulsion, the equilibration concentration profiles depicted in figure~\ref{Fig3_oil_in_water_And_Species} were utilitzed for electrochemical simulations under Butler--Volmer electrode kinetics. 
In case of the equipartition depicted in panels I a) and I b) of figure~\ref{Fig3_oil_in_water_And_Species}, i.e. a \textit{'micro-emulsion'} which is almost entirely a non-polar phase, it was expected that --- a) if the first reaction is fast and b) the second reaction is very sluggish --- the classical (Randles--\v{S}ev\v{c}ík like) CV will be obtained. Indeed, an excellent agreement between the numerically computed CV based on the theory presented in this paper with the semi-analytical result simulated with the free software tool~\textit{Polarographica}~\cite{Tichter2019b, Polarograhica2021} (dotted curve) is obtained, which can be seen in figure~\ref{Fig4_Classical_Comparison}. Furthermore, figure~\ref{Fig4_Classical_Comparison} depicts the concentration profiles of all active species which are essentially only $A_{\textrm{o}}$ and $B_{\textrm{o}}$. Parameters used in the simulation are stated along with the figure caption.

\begin{figure}[H] 
\begin{center}
\includegraphics[width=11cm, height=14.5cm]{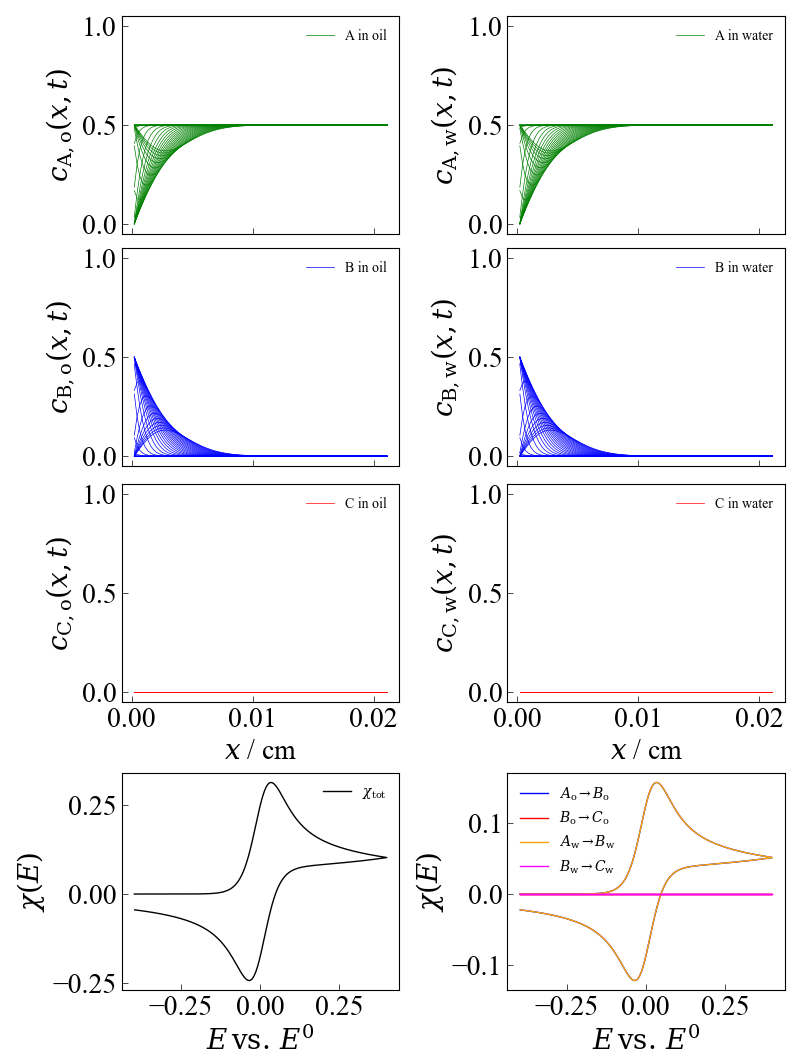}
\caption{CV curves simulated for the spatial equipartition of non-polar to polar phases in 1/1, depicted in figure~\ref{Fig3_oil_in_water_And_Species} II a) and I b). For the simulation it was set $E^{0}_{1} = E^{0}_{3} = 0.00\,\textrm{V}$, $E^{0}_{2} = E^{0}_{4} = 0.05\,\textrm{V}$, $\alpha_{1} = \alpha_{2} = \alpha_{3} = \alpha_{4} = 0.5$, $k^{0}_{1} = k^{0}_{3} = 10^{3}\,\textrm{cm/s}$, $k^{0}_{2} = k^{0}_{4} = 10^{-23}\,\textrm{cm/s}$, $D_{A,o}^{0} = D_{B,o}^{0} = D_{C,o}^{0} = D_{A,w}^{0} = D_{B,w}^{0} = D_{C,w}^{0} = 10^{-6}\,\textrm{cm}^{2}\textrm{/s} $, $A = 1\,\textrm{cm}^{2}$, $c_{\textrm{tot}} = 1\,\textrm{mol/L}$, $k_{1} = 5.5~\textrm{mol/Ls}$, $k_{-1} = 5.5\,10^{-1}~\textrm{mol/Ls}$, $k_{2} = 10^{-3}~\textrm{mol/Ls}$, $k_{3} = 10^{-3}~\textrm{mol/Ls}$ ($k_{-2}$ and $k_{-3}$ are thus defined from the thermodynamics). The function $o(x)$ was defined by setting $\mu = 50$ and $x_{\textrm{T}} = 0.1\,x_{\textrm{max}}$, $o_{\textrm{rel}}(\infty) = 0.5$, $o_{\textrm{rel}}(0) = 0.5$. The total concentration of the polar phase was set to $55\,\textrm{mol/L}$ and the concentration of the non-polar phase to $5\,\textrm{mol/L}$.}
\label{Fig5_TwoPhaseEqDist}
\end{center}
\end{figure}

In contrast to figure~\ref{Fig4_Classical_Comparison}, figure~\ref{Fig5_TwoPhaseEqDist} shows the CV curves for the spatial distribution of the polar and non-polar phases and the resulting profiles of electrochemically active species depicted in figure~\ref{Fig3_oil_in_water_And_Species} II a) and II b). The homogeneous rate constants were adjusted in a way that they compensate the ratio in the total concentration of polar and non-polar phases such that $A_{\textrm{o}}$ and $A_{\textrm{w}}$ are initially present in the same amount.   All dimension-less CVs possess a peak height reduced by the factor $\sqrt{2}$ when compared to the depicted in terms of the dimension-less flux, since they were normalized to $\sqrt{D_{\textrm{A,o}}^{0}}$. The effective diffusion coefficient would be, however, $\sqrt{0.5\,D_{\textrm{A,o}}^{0}}$ owing to the 1/1 ratio of polar and non-polar phase. Since the kinetics from the oil and water phase were assumed to be equal, the individual (dimension-less) currents from the reaction $A_{\textrm{o}} \to B_{\textrm{o}}$ and $A_{\textrm{w}} \to B_{\textrm{w}}$ are identical.

\begin{figure}[H] 
\begin{center}
\includegraphics[width=11cm, height=14.5cm]{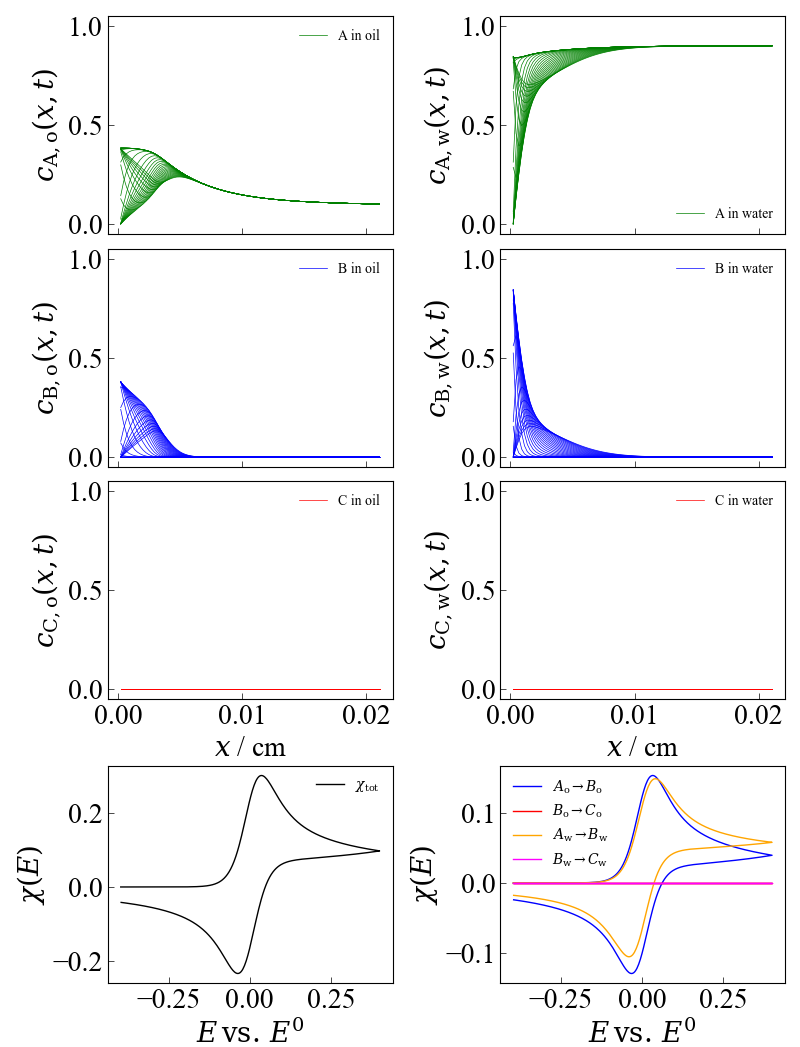}
\caption{CV curves simulated for the spatial equipartition of non-polar to polar phases in 1/1, depicted in figure~\ref{Fig3_oil_in_water_And_Species} II a) and I b). For the simulation it was set $E^{0}_{1} = E^{0}_{3} = 0.00\,\textrm{V}$, $E^{0}_{2} = E^{0}_{4} = 0.05\,\textrm{V}$, $\alpha_{1} = \alpha_{2} = \alpha_{3} = \alpha_{4} = 0.5$, $k^{0}_{1} = k^{0}_{3} = 10^{3}\,\textrm{cm/s}$, $k^{0}_{2} = k^{0}_{4} = 10^{-23}\,\textrm{cm/s}$, $D_{A,o}^{0} = D_{B,o}^{0} = D_{C,o}^{0} = D_{A,w}^{0} = D_{B,w}^{0} = D_{C,w}^{0} = 10^{-6}\,\textrm{cm}^{2}\textrm{/s} $, $A = 1\,\textrm{cm}^{2}$, $c_{\textrm{tot}} = 1\,\textrm{mol/L}$, $k_{1} = 5.5~\textrm{mol/Ls}$, $k_{-1} = 5.5\,10^{-1}~\textrm{mol/Ls}$, $k_{2} = 10^{-3}~\textrm{mol/Ls}$, $k_{3} = 10^{-3}~\textrm{mol/Ls}$ ($k_{-2}$ and $k_{-3}$ are thus defined from the thermodynamics). The function $o(x)$ was defined by setting $\mu = 50$ and $x_{\textrm{T}} = 0.1\,x_{\textrm{max}}$, $o_{\textrm{rel}}(\infty) = 0.1$, $o_{\textrm{rel}}(0) = 0.85$. The total concentration of the polar phase was set to $55\,\textrm{mol/L}$ and the concentration of the non-polar phase to $5\,\textrm{mol/L}$.}
\label{Fig6_TwoPhaseEqDist_unequal}
\end{center}
\end{figure}

Figure~\ref{Fig6_TwoPhaseEqDist_unequal} depicts the effect of the anisotropic distribution of polar and non-polar phases shown in figure~\ref{Fig3_oil_in_water_And_Species} III a) and III c) on the electrochemistry of the system. It can be seen that in the concentration profiles in the non-polar phase, a significant kink emerges within time. This feateure is ascribed to the spatially dependent diffusion coefficients and can be understood as follows: The local enrichment of non-polar solvent in front of the electrode facilitates the mass transfer of the non-polar species in direct proximity to the reactive surface. The species formed by the electrochemical reaction is, however, merely restricted to this particular \textit{'oil-layer'}. Likewise, diffusion to and from the bulk of the electrolyte is significantly hindered. As a consequence, the entire mass transfer tends towards the scenario of a thin-layer cell, i.e. it becomes similar --- but not equal --- to a finite reflective diffusion. This can be seen also in the individual (dimension-less) CV response from the reaction $A_{\textrm{o}} \to B_{\textrm{o}}$. Whereas the reaction $A_{\textrm{w}} \to B_{\textrm{w}}$ possesses more or less the classical diffusion tail, the decaying part of the CV of the reaction $A_{\textrm{o}} \to B_{\textrm{o}}$ is significantly steeper. Additionally, the \textit{\textquotesingle back-peak\textquotesingle} is more pronounced, since the species formed at the electrode surface cannot diffuse too far into the electrolyte as it was the case in an isotropic electrolyte. 

\begin{figure}[H] 
\begin{center}
\includegraphics[width=11cm, height=14.5cm]{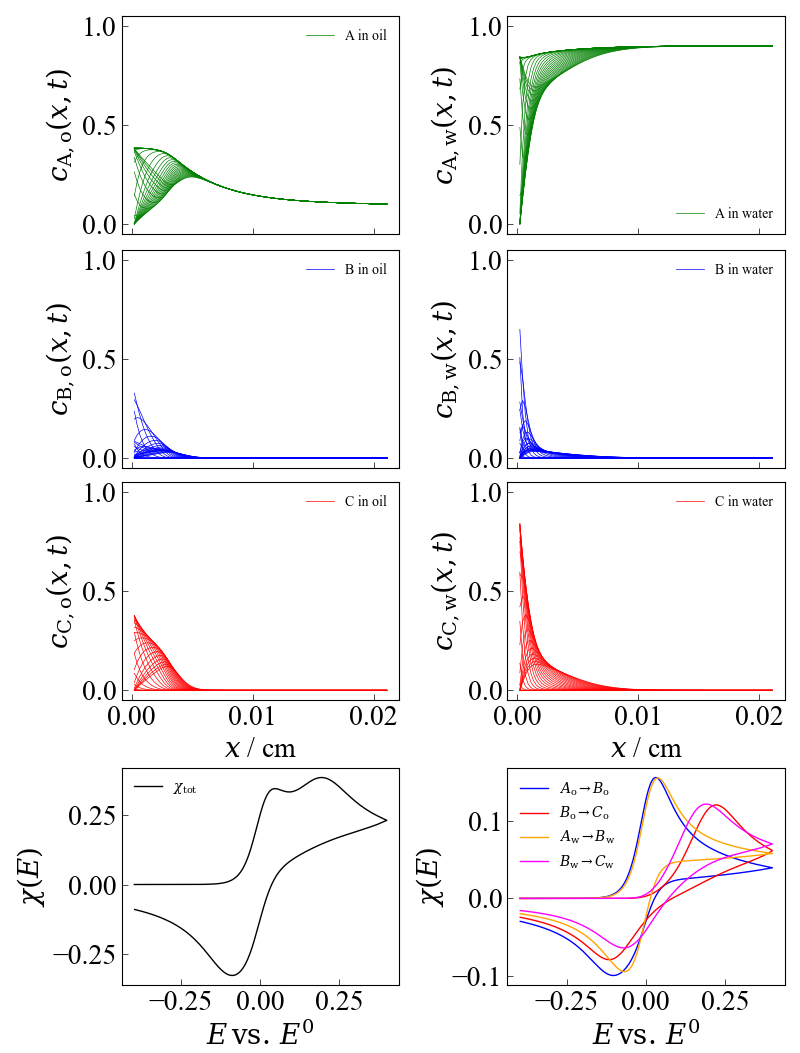}
\caption{CV curves simulated for the spatial equipartition of non-polar to polar phases in 1/1, depicted in figure~\ref{Fig3_oil_in_water_And_Species} II a) and I b). For the simulation it was set $E^{0}_{1} = E^{0}_{3} = 0.00\,\textrm{V}$, $E^{0}_{2} = E^{0}_{4} = 0.05\,\textrm{V}$, $\alpha_{1} = \alpha_{2} = \alpha_{3} = \alpha_{4} = 0.5$, $k^{0}_{1} = k^{0}_{3} = 10^{3}\,\textrm{cm/s}$, $k^{0}_{2} = k^{0}_{4} = 10^{-4}\,\textrm{cm/s}$, $D_{A,o}^{0} = D_{B,o}^{0} = D_{C,o}^{0} = D_{A,w}^{0} = D_{B,w}^{0} = D_{C,w}^{0} = 10^{-6}\,\textrm{cm}^{2}\textrm{/s} $, $A = 1\,\textrm{cm}^{2}$, $c_{\textrm{tot}} = 1\,\textrm{mol/L}$, $k_{1} = 5.5~\textrm{mol/Ls}$, $k_{-1} = 5.5\,10^{-1}~\textrm{mol/Ls}$, $k_{2} = 10^{-3}~\textrm{mol/Ls}$, $k_{3} = 10^{-3}~\textrm{mol/Ls}$ ($k_{-2}$ and $k_{-3}$ are thus defined from the thermodynamics). The function $o(x)$ was defined by setting $\mu = 50$ and $x_{\textrm{T}} = 0.1\,x_{\textrm{max}}$, $o_{\textrm{rel}}(\infty) = 0.1$, $o_{\textrm{rel}}(0) = 0.85$. The total concentration of the polar phase was set to $55\,\textrm{mol/L}$ and the concentration of the non-polar phase to $5\,\textrm{mol/L}$.}
\label{Fig7_TwoPhaseEqDist_unequal_EkEk}
\end{center}
\end{figure}

Figures~\ref{Fig4_Classical_Comparison} to~\ref{Fig6_TwoPhaseEqDist_unequal} have in common, that they exclude a second electrochemical reaction (i.e. they consider an E\textsubscript{K} instead of an E\textsubscript{K}E\textsubscript{K} reaction in the square scheme) since the heterogeneous rate constants of the second step were set to $k^{0}_{2} = k^{0}_{4} = 10^{-23}~\textrm{cm/s}$. To illustrate the influence of a second electron transfer which can occur from both phases, figure~\ref{Fig7_TwoPhaseEqDist_unequal_EkEk} depicts the scenario with a quasireversible second step by setting $k^{0}_{2} = k^{0}_{4} = 10^{-4}~\textrm{cm/s}$. This leads, of course, to four individual current responses, i.e. one per reaction and phase as $A_{\textrm{o}} \to B_{\textrm{o}}$, $B_{\textrm{o}} \to C_{\textrm{o}}$ and $A_{\textrm{w}} \to B_{\textrm{w}}$ and $B_{\textrm{w}} \to C_{\textrm{w}}$. The total current possesses a split \textit{\textquotesingle forward peak\textquotesingle} and a single \textit{\textquotesingle back-peak\textquotesingle} in this case, which can be understood readily by considering the sum of the individual currents. Since the second electron transfer reactions are set to occur $+50~\textrm{mV}$ beyond $E^{0}_{1}$ and $E^{0}_{3}$ (i.e. at $E^{0}_{2} = E^{0}_{4} = 0.05\,\textrm{V}$) and since the $k^{0}_{2} = k^{0}_{4}$ introduce rather sluggish electrode kinetics, the \textit{\textquotesingle forward peaks\textquotesingle} are well separated. However, the \textit{\textquotesingle backward-peaks} of the individual currents overlap (by coincidence of the simulation --- or experiment) in this case since the sluggish electrode kinetics enforce a stronger peak-to-peak separation. This underlines that a double \textit{\textquotesingle forward peak\textquotesingle} and a single \textit{\textquotesingle back-peak\textquotesingle} does not mean that one of the forward reactions cannot be reversed. By regarding the concentration profiles in figure~\ref{Fig7_TwoPhaseEqDist_unequal_EkEk}, it can be seen that species $B$ is now consumed by a follow up reaction. Consequently, the respective concentration profiles are much less pronounced. Of course, there is also a contribution in the concentration profiles of species $C$, which is formed from $B$ in this case. 

For an animated visualization of the time-dependent concentration profiles of figures~\ref{Fig4_Classical_Comparison} to~\ref{Fig7_TwoPhaseEqDist_unequal_EkEk}, the reader is herewith referred to the video material in the supporting information of this paper.

\subsection{Theory versus Experiment}

\begin{figure}[H] 
\begin{center}
\includegraphics[width=11cm, height=9.07cm]{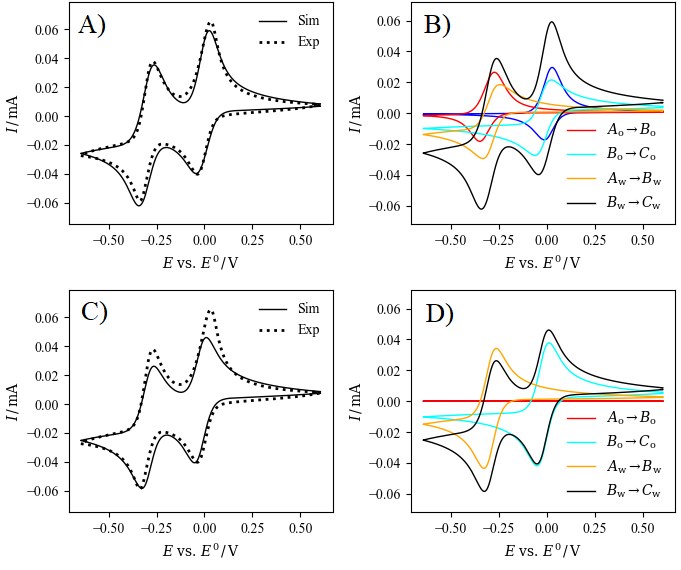}
\caption{Experimentally acquired (dotted curves) and simulated (solid traces) data for: A) and B), a micro-emulsion with an anisotropic --- and C) and D) a micro-emulsion with an isotropic distribution of polar and non-polar phases. It can be seen that the model of an anisotropic ME reproduces the experimentally observed CV much more accurate. The following set of parameters was found to qualitatively reproduce the experimentally acquired data exceptionally well. $E^{0}_{2}-E^{0}_{1} = 0.295\,\textrm{V}$, $E^{0}_{3}-E^{0}_{1} = 0.025\,\textrm{V}$, $E^{0}_{4} - E^{0}_{1} = 0.295\,\textrm{V}$, $\alpha_{1} = \alpha_{2} = \alpha_{3} = \alpha_{4} = 0.5$, \linebreak $k^{0}_{1} = k^{0}_{3} = k^{0}_{4} = 15\times 10^{-3}\,\textrm{cm/s}$, $k^{0}_{2} = 1\times 10^{-3}\,\textrm{cm/s}$, $D_{A,o}^{0} = 3.0\times 10^{-6}\,\textrm{cm}^{2}\textrm{/s}$, $D_{B,o}^{0} = 7.5\times 10^{-6}\,\textrm{cm}^{2}\textrm{/s}$, $ D_{C,o}^{0} = 2.0\times 10^{-6}\,\textrm{cm}^{2}\textrm{/s}$, $D_{A,w}^{0} = 6.2\times 10^{-6}\,\textrm{cm}^{2}\textrm{/s}$, \linebreak $D_{B,w}^{0} = 9.0\times 10^{-6}\,\textrm{cm}^{2}\textrm{/s}$, $D_{C,w}^{0} = 8.0\times 10^{-6}\,\textrm{cm}^{2}\textrm{/s} $, $c_{\textrm{tot}} = 10\,\textrm{mmol/L}$, $k_{1} = 2.00~\textrm{mol/Ls}$, $k_{-1} = 0.25~\textrm{mol/Ls}$, $k_{2} = 5~\textrm{mol/Ls}$, $k_{3} = 150~\textrm{mol/Ls}$. For the anisotropic ME, the function $o(x)$ was defined by setting $\mu = 350$ and $x_{\textrm{T}} = 3\,\mu\textrm{m}$, $o_{\textrm{rel}}(\infty) = 0.0133$, $o_{\textrm{rel}}(0) = 0.85$. The total concentration of the polar phase was set to $40.13\,\textrm{mol/L}$ (resulting from the concentration of water in the phase surfactant/water) and the total concentration of the non-polar phase to $9.44\,\textrm{mol/L}$ (i.e. the molar concentration of toluene). For the isotropic ME, the only change was to set $o_{\textrm{rel}}(0) = 0.0133$. The potential sweep rate was set to $\nu = 10~\textrm{mV/s}$ (according to the experiment). The numerical resolution of the simulation was set to $\Delta \xi = 0.01$.}
\label{Fig8_Fitting}
\end{center}
\end{figure}

To support our theoretical model for the electrochemistry in anisotropic micro-emulsions, our simulations are coroborated by experimental data acquired for the electrochemical two-step redox reaction of methyl-viologen (MV) in a ME of toluene in water (cf. experimental section in the SI). In figure~\ref{Fig8_Fitting} A) and C) the experimentally measured redox-current is shown as dotted curve. The result from the simulation is depicted as a solid trace. Panels B) and D) of figure~\ref{Fig8_Fitting} depict the individual current contributions from the redox reactions of the active species in the non-polar and polar phases, respectively. 

For the simulation of panels A) and B), a spatially anisotropic ME was considered. In all simulations it was assumed that --- according to the experiment --- the ME had a relative content of 1.33wt.\% toluene, 72wt.\% water and 26.67wt.\% surfactant (the latter two quantities are regarded as the polar phase) in the bulk. An enrichment of 85\% oil concentration in front of the electrode with an $x_{\textrm{T}} = 3~10^{-6}~\textrm{m}$ and a parameter $\mu =350$ was found to accurately reproduce the experimentally measured data. In contrast to panels A) and B), panels C) and D), were base on the assumption of an equipartition of non-polar and polar phases. It can be seen that the CV curves computed on the base of this isotropic diffusion model provides a significantly worse fit of the experimental data in the \textit{\textquotesingle back-peaks\textquotesingle} which can be explained as follows. 

In case of the isotropic ME (a spatially independent ratio of toluene) the forward reactions (reduction of $MV^{2+}\to MV^{+} \to MV$) take place merely from the aqueous phase. This leads to the classical semi-infinite diffusion which fits the \textit{\textquotesingle forward-peaks\textquotesingle} of the experimentally measured data fairly accurate. Nevertheless, since the concentration of toluene in direct proximity to the electrode is comparably low, there is only a minor transfer of $MV^{+}$ and $MV$ to the non-polar phase (precipitation excluded here). Consequently, the reverse reaction (here the oxidation) can only follow the characteristics of a semi-infinite diffusion (i.e. from the aquous phase only) as well. Likewise, the CV curves will only show the classical semi-infinite diffusion characteristics, i.e. a rather Randles-\v{S}ev\v{c}ík-like smooth tailing.

\newpage

In case of the anisotropic diffusion, i.e. an enrichment of the non-polar phase in front of the electrode, the \textit{\textquotesingle back-peaks\textquotesingle} will behave more or less like a finite-diffusion CV (i.e. they will become more pronounced and have a steeper decay). This is mainly caused by the fact a significant proportion of the current can be drawn from the non-polar phase and that the concentrations of $MV^{+}$ and $MV$ will be locally enriched (or trapped in the oil layer) by the coupled chemical reactions. In this scenario, the \textit{\textquotesingle pseudo-finiteness\textquotesingle} of the diffusion domain is introduced by the fact that locally enrichment in the oil phase is present over a few micrometers only and the effective diffusion will be slowed down significantly in the bulk of the electrolyte. Since such an anisotropy in the diffusion domain is a straighforward --- and figurative --- way of reproducing experimentally measured CV data, we regard it as the final validation of our theory.

\section{Summary and Conclusions}
In the present paper we have derived the theory of cyclic voltammetry for a kinetically controlled two-electron reaction  which takes place in a spatially anisotropic micro-emulsion. By re-deriving the diffusion equation to account for spatially  dependent diffusion coefficients we capture the complex mass transfer phenomena in a ME, where a local enrichment of a non-polar phase is present in front of the reactive electrode surface. In this manner, we are able to accurately reproduce experimentally acquired data for the two-electron redox reaction of methyl-viologen in a micro-emulsion of toluene in water , which is not possible by considering an isotropic ME. Consequently, we point out that the model presented in this paper paves the way for decent investigations of the electrochemistry of micro-emultions --- an analytical tool which is of utmost interest in the context of organic redox-flow batteries.
\newpage

\bibliographystyle{angew}

\bibliography{\jobname}
\end{document}